\documentclass[final,3p,times]{elsarticle}

\usepackage{amssymb}
\usepackage{multirow}

\journal{Transportation Research Part D: Transport and Environment}

\begin{document}

\begin{frontmatter}



\title{Can autonomy make bicycle-sharing systems more sustainable? Environmental impact analysis of an emerging mobility technology}


\author[inst1]{Naroa Coretti Sanchez}
\author[inst1]{Luis Alonso Pastor}
\author[inst1]{Kent Larson}

\affiliation[inst1]{organization={MIT Media Lab},
            addressline={20 Ames Street}, 
            city={Cambridge},
            postcode={02142}, 
            state={Massachusetts},
            country={United States}}



\begin{abstract}

Autonomous bicycles have recently been proposed as a new and more efficient approach to bicycle-sharing systems (BSS), but the corresponding environmental implications remain unresearched. Conducting environmental impact assessments at an early technological stage is critical to influencing the design and, ultimately, environmental impacts of a system. Consequently, this paper aims to assess the environmental impact of autonomous shared bikes compared with current station-based and dockless systems under different sets of modeling hypotheses and mode-shift scenarios. The results indicate that autonomy could reduce the environmental impact per passenger kilometer traveled of current station-based and dockless BSS by 33.1 \% and 58.0 \%. The sensitivity analysis shows that the environmental impact of autonomous shared bicycles will mainly depend on vehicle usage rates and the need for infrastructure. Finally, this study highlights the importance of targeting the mode replacement from more polluting modes, especially as traditional mobility modes decarbonize and become more efficient.

\end{abstract}


\end{frontmatter}

\section{Introduction}

By 2018 55\% of the world's population lived in cities, and this number is expected to grow to 68\% by 2050 \cite{UNreport2018}. Combined with the effect of population growth, this process would add 2.5 billion people to urban areas by 2050, posing a challenge for sustainable development \cite{UNreport2018, UNreport2019}. Transportation, in particular, is one of the leading sustainability challenges in cities. For instance, in the United States, the transportation sector is the most significant contributor to greenhouse gas (GHG) emissions. In 2019, it accounted for 29 \% of the emissions, most of it coming from light-duty vehicles, which include passenger vehicles and pickup trucks \cite{EPA2021}. Despite the measures taken to reduce the environmental impact of mobility, transportation-related emissions are still growing, and the total vehicle miles traveled are likely to keep increasing \cite{EPA2021,leard2019explaining}. This predicted evolution indicates the urgency of finding mobility solutions that provide an efficient and ecological service in urban areas. 
 
Electrification, vehicle-sharing, and autonomy the main trends predicted to transform urban mobility in the coming years \cite{McKinsey2019}. While an appropriate combination of these trends could have a positive environmental impact, existing studies highlight the importance of reducing vehicle miles traveled in order to meet carbon emission targets \cite{alarfaj2020decarbonizing}. One of the most popular actions taken by cities to promote active mobility modes and reduce vehicle travel has been to implement bicycle-sharing systems (BSS). As a consequence, since 2005 the number of station-based BSS worldwide grew from 17 systems to 2147 in 2019 \cite{sanchez2020autonomous}. More recently, other micro-mobility systems such as dockless bikes and e-scooters have become a mainstream inner-city transportation mode. For instance, in the US, the total annual micro-mobility trips have drastically increased from 35M in 2017 to 136M in 2019, and this growth rate has further accelerated as a consequence of the Covid-19 pandemic \cite{nacto2019,reck2021mode}.

However, as BSS became increasingly popular, cities started to witness some of their challenges. One of the most relevant challenges is the bicycle rebalancing problem, which is caused by user travel patterns being unbalanced in space and time \cite{ScientometricReview, curran2008,lin2012geo}. To mitigate this issue, operators transport bicycles from saturated areas in vans or trucks, a process that has a very high economic and ecological cost \cite{OBIS2011, luo2019comparative}. For the users, rebalancing problem often leads to difficulties in finding available bikes and docks, increasing travel times, and leading to frustration and loss of reliability in the system \cite{Kaltenbrunner2010, nair2013, lin2012geo}.
Another very relevant challenge is related to the bicycle oversupply in dockless systems. Not being restricted by the number of docks, dockless BSS operators have followed a "winner takes all" approach, flooding cities with dockless fleets up to 10 to 60 times larger than the existing station-based systems \cite{tao2021evaluation, DocklessBS}. These fleet sizes vastly exceed user demand and the infrastructure capacities of cities, provoking system inefficiency and urban problems such as bicycles being parked illegally clogging sidewalks \cite{ScientometricReview,zhao2020bike}. As a consequence of these problems, many cities have limited the fleet sizes or forced operators to cease their programs \cite{ shi2018}. Together with the saturation of the market, this has caused many companies to go bankrupt, and the amount of discarded bicycles has generated an enormous recycling crisis \cite{DocklessBS, shi2018}. 

As a consequence of these issues, around 2019, scholars started to raise concerns about the lack of studies related to the sustainability of BSS, which was especially alarming considering that it had been almost a decade since their popularization. Since then, there has been an increasing number of studies analyzing their environmental impact \cite{luo2019comparative, luo2020optimizing, zheng2019bicycle, de2021environmental,bonilla2020life,d2021emissions, sun2020environmental, mao2021can, sun2021contribution,chen2020life}. One of the early assessments on the environmental impacts of bicycle-sharing was carried out by Luo et al. \cite{luo2019comparative} who conducted a Life Cycle Assessment (LCA) of station-based and dockless bicycles. They later extended their work to understanding the environmental trade-offs between fleet sizing and rebalancing, which is one of the critical decisions in designing BSS \cite{luo2020optimizing}. Another factor that heavily influences the impact of a BSS is the share of mobility modes displaced by bike-sharing trips; therefore, other studies have focused on mode shift assessment \cite{reck2021mode,li2021high}. 

Recently, the introduction of autonomous driving technology into bicycle-sharing systems has been proposed as a solution that would combine the most relevant benefits of vehicle sharing, electrification, autonomy, and micro-mobility to solve some of the challenges currently faced with BSS \cite{sanchez2020autonomous,lin2021affordable}. The premise of these systems is that, since autonomous bicycles would be able to relocate by themselves, autonomous BSS would work as a mobility-on-demand system - similar to Uber or Lyft - eliminating the problem of users needing to find available bikes or docks. At the same time, eliminating the need for rebalancing could potentially lead to cost savings for system operators \cite{sanchez2020autonomous}. From a technological readiness perspective, while autonomous bicycles are at an early technology stage, there have been numerous studies ranging from vehicle design to control or computing systems related to autonomous micro-mobility systems such as bicycles, tricycles, or scooters \cite{kamga1996speed,andersen2016autonomous,pandey2017modeling,luo2018porca,mulky2018autonomous,chen2020designing,wenzelburger2020first,yu2020building, christensen2021autonomous, kondor2021estimating}. In addition, our research group has developed autonomous micro-mobility systems for over a decade, including the development of an ultra-lightweight autonomous tricycle project named the Persuasive Electric Vehicle (PEV) \cite{lin2021affordable,yao2019idk,agarwal2018design}. These projects had already proved the viability of autonomous tricycles; therefore, we focused in addressed the main challenge of translating this technology to a bicycle: lateral stability. For that, we designed, prototyped, and tested a vehicle that dynamically transforms from a bicycle configuration into a tricycle configuration for the autonomous drive \cite{sanchez2020autonomous}. We then analyzed the fleet behavior of autonomous bicycle-sharing systems in a previous simulation study, quantifying the extent to which it would outperform current bicycle-sharing schemes \cite{sanchez2021simulation}. The results showed that an autonomous BSS could improve performance and user experience with a fleet size three and a half times smaller than a station-based system and eight times smaller than a dockless system. Since bicycle manufacturing and rebalancing are some of the main drivers for the environmental impacts of bike-share \cite{luo2019comparative}, having fewer bikes and eliminating the need for rebalancing could potentially reduce the environmental impact of BSS. 

This study builds on our previous work to compare autonomous and current station-based and dockless BSS from an environmental perspective. Conducting environmental impacts assessments of technology at such an early technology development stage presents a significant challenge due to real-world data and associated uncertainty. However, it is also well-known that it is critical to understand its possible implications while there is a great opportunity to influence their design and, ultimately, its environmental impacts \cite{collingridge1982social,bergerson2020life}. According to the life cycle assessment (LCA) methodology for emerging technologies proposed by Bergerson et al. \cite{bergerson2020life}, an autonomous BSS fits in what they define as an 'Emerging technology/ Emerging market' scenario, which is the one with the largest uncertainty. Bergerson et al. suggest that, in this scenario, simplified LCAs are more appropriate and that efforts should be focused on exploring a broader set of scenarios and performing sensitivity analyses that help identify the key parameters that impact the environmental performance \cite{bergerson2020life,lacirignola2017lca}.

Following these recommendations, this study takes a simplified LCA approach to compare the life-cycle environmental impacts of autonomous and current station-based and dockless systems focusing on broader scenario analysis and parameter variations. Even if LCA is a valuable tool for understanding the environmental impacts of products and services, the results are heavily influenced by methodological choices \cite{ITF}. Therefore, even if we have performed sensitivity analyses to quantify the impacts of the modeling hypotheses, we advise readers to keep in mind that environmental impacts are very case-specific \cite{bergerson2020life, ITF}.

\section{Life Cycle Assessment} \label{lca} 

The evaluation of the environmental impacts of autonomous bike-share versus current station-based and dockless systems has been based on a comparative life cycle assessment (LCA). LCA is a standardized environmental impact calculation method that provides a quantitative analysis of the impacts of a service or product across the entire life cycle, from raw material acquisition, via production and use phases, to waste management \cite{ISO, finnveden2009recent}. 

As mentioned above, conducting an LCA for an emerging technology represents an additional challenge due to the uncertainties associated with the lack of real-world data \cite{collingridge1982social,bergerson2020life}. Therefore, we have followed the methodology proposed by Bergerson et al. for these scenarios, which consists of a simplified LCA focused on parameter variation and scenario analysis \cite{bergerson2020life}. The specific methodology followed was the LCA methodology adaptation for new mobility proposed by the International Transport Forum (ITF) \cite{ITF}. The reason for choosing this methodology is that it adds a new component to account for operational services that are characteristic of new mobility systems and are one of the main contributors to their environmental impact \cite{ITF,luo2019comparative}. Alongside this methodology, the ITF published an assessment tool specifically designed to estimate the environmental impacts of new mobility. This tool models traditional transportation systems and applies their adapted LCA methodology to emerging mobility modes. The flexibility of this tool has allowed us calculate the environmental impact of an autonomous BSS, modify their modeling assumptions for station-based and dockless systems based recent literature, and perform a sensitivity analysis of our hypotheses.

An LCA comprises four main stages: the definition of the goal and scope, Life Cycle Inventory Analysis (LCI), Life Cycle Impact Assessment (LCIA), and interpretation \cite{ISO}. These stages will be covered in the following sections.

\subsection{Goal and scope}

The goal of this study is threefold: First, we estimate the comparative life cycle environmental impact of autonomous, station-based and dockless electric BSS. Second, we identify the key parameters affecting the environmental impact of autonomous BSS by studying the impacts of variations of the main modeling hypotheses. Third, we estimate the results' implications when considering the mode shift generated by bike-share trips.

As the goal of this study is to assess the environmental impact of the movement of people, the functional unit used in this study is one passenger kilometer traveled (pkm). In terms of scope, scope of our LCA is defined by the following components as defined in the methodology by \cite{ITF}:  
\begin{itemize}
\item A vehicle component, considering manufacturing, delivery, maintenance, and end-of-life of the vehicle and its components.
\item A fuel component, considering both the well-to-tank and tank-to-wheel phases.
\item An infrastructure component, considering the materials extraction, processing, construction, maintenance, and end of life. A key difference in this study is that the impact of the docking and charging stations have also been included within the system's boundaries, which, as will be discussed in Section \ref{results}, has a significant impact on lifecycle emissions.
\item An operational component that expands traditional transport LCAs to account for the servicing need, which is characteristic of new mobility. This component includes servicing operations such as recharging, rebalancing, or passenger pick-up (i.e., overheading). A well-to-wheel phase is considered for the servicing vehicles, assuming that these vehicles were not manufactured solely for this purpose and are used for other purposes \cite{ITF}. For example, for station-based bicycles, the operational service is bicycle rebalancing. In the case of the dockless system, instead, bicycles need rebalancing and transportation for charging. Lastly, in the autonomous system, servicing includes autonomous driving trips for charging and overheading.
\end{itemize}

\subsection{Inventory analysis}\label{inventory}

The inventory analysis has been based on available literature complemented with our research lab's research studies and prototype data. Since this study aims to provide a comparative assessment of the impacts of different bike-sharing systems, the demand has been considered constant and we have applied a consistent set of assumptions for all the systems. The hypotheses for the nominal case of our study are specified in this section, and they have been accordingly reflected in the different steps of the LCA: manufacturing, transport, use, operational services, and infrastructure. The assumptions related to the inventory analysis are outlined below.

\subsubsection{Vehicle lifetime}

The Institute for Transportation \& Development Policy (ITDP) bike-share planning guide suggests that shared bikes should be designed to have a lifespan of 3-5 years \cite{ITDP}. However, shared-bike service industry standards in cities like Beijing and Shanghai specify that the bikes should be scrapped after three years, even if the current service life can be longer \cite{tao2021evaluation, Sun2021contr,sun2021does}. Therefore, we consider the average lifetime of each bicycle to be three years. 

\subsubsection{Vehicle utilization}

To calculate the annual mileage, the average mileage of each trip is based on \cite{ITF} and the daily vehicle utilization rates are based on our previous simulation study that compared the performance of station-based, dockless and autonomous systems serving the same demand \cite{sanchez2021simulation} and . In the case of the autonomous system, the annual mileage must also include the distance that the bicycles travel for pick-up and charging. Since the original study intended to evaluate the impact of autonomy, it did not consider the electrification of the user trip. Therefore,  the simulation tool has been adapted for this study to include battery consumption during the ride. The results obtained reflect an overhead distance of 21.71 \% and a charging distance of 0.43 \%. Having data from the three systems under the same demand has been critical to this study since it helps avoid errors from comparing their performance under different scenarios. . 

\subsubsection{Vehicles}

The weight of the autonomous bicycle has been considered to as an average between the weight of an e-bike (30.8 kg) and the PEV (50 kg) \cite{ITF, PEV,lin2021affordable}. The battery capacity of the autonomous bicycle has also been set to be the same as the battery of the PEV (0.4896 kWh) \cite{lin2021affordable}. The electronics needed to support autonomous driving have been considered to be the same as for the PEV: two computers, two cameras, two small LiDAR-s, one sonar, one radar, one GPS unit, and a UWB system \cite{lin2021affordable}. The life-cycle environmental impact of these components has been based on the work of Gawron et al. \cite{gawron2018life}, selecting the corresponding components and adjusting the use phase emissions considering the lifetime mileage of autonomous bicycles. The environmental impact of the UWB system has been considered to be the same as the DSRC system reported in \cite{gawron2018life}. These components added a total of 599.3 kg$\mathrm{CO_2}$ eq to the manufacturing impact of the autonomous bicycles. However, in a future in which autonomous vehicles are widespread in the urban environment, cities will have infrastructure that will support autonomy minimizing the intelligence and computational power requirements of each vehicle. If the technology supporting the autonomy was shared, the related environmental impact could be significantly reduced. Lastly, the impact of the station-based and dockless e-bikes has been extracted from \cite{ITF}. 

\subsubsection{Rebalancing}

For the dockless system, the need for rebalancing in terms of service vehicle kilometers per passenger kilometer traveled has been considered to be 100 m/pkt as reported by \cite{ITF}. With this assumption, we consider that the need for rebalancing is determined by the overall system demand instead of being proportional to the number of vehicles in the system. Having e-bikes instead of regular bikes does not impact rebalancing in the station-based system because bikes get charged at docking stations. Therefore, the need for rebalancing vehicle kilometers per passenger kilometer traveled has been assumed to be the same as in Luo et al. \cite{luo2019comparative}. 

\subsubsection{Power mix}

Since the vehicle usage is based on our previous simulation study based on the Metro  Boston  Region \cite{sanchez2021simulation} , for the power generation, we have considered the electricity mix from the United States. As reported by the International Energy Agency (IEA) \cite{IEA} the electricity mix in  the United States for 2020 was the following one: 39 \% natural gas, 20 \% coal, 19 \% nuclear, 19\% renewable, 1 \% oil and 1 \% biomass.

\subsubsection{Infrastructure}
The impact of charging stations has been extracted from an average of the two scenarios presented by Mondello et al. \cite{mondello2018environmental}. These results have been adapted for this study by removing the impacts from manufacturing eight e-bikes and the use phase since we had already considered these impacts in the manufacturing and general use phases, respectively. The number of stations per bike-km in the station-based system has been considered to be 4.28e-06 station/bike-km as in Coretti et al. \cite{sanchez2021simulation} which is based on Boston's Bluebikes system. The stations in the autonomous system have been considered twice the one in a station-based system on a per bicycle basis as in\cite{sanchez2020autonomous}, leading to a value of 2.5e-06 station/bike-km. Due to lack of data, the dockless system has been considered to have the same total infrastructure impact as the station-based system because, even if dockless BSS do not need docking stations, the fleets are significantly more extensive, and there is a need for charging depots. In terms of road usage, we consider that bikes travel 80 \% of the miles on bike lanes and 20 \% of the time on urban roads and include the additional weight of the autonomous bike into the modulation of road usage share.

\subsection{Impact assessment} \label{impact-assessment}

The life cycle impact assessment has been performed by integrating the inventory analysis described in Section \ref{inventory} into ITF's assessment tool, developed to estimate the environmental impacts of new mobility \cite{ITF}. One of the main advantages of this procedure is that it applies a consistent set of assumptions on the carbon intensity of materials and energy sources for all the modes analyzed, including the mobility modes displaced by bike-sharing, as shown in Section \ref{mode-substitution}. Our analysis focuses on GHG emission impacts because transportation is a major source of $\mathrm{CO_2}$ emissions and has had a historical dependence on fossil fuels and GHG emissions play a central role in economic and political debates and climate action plans of cities \cite{EPA2021}. However, a more complete LCA analysis should be conducted to understand the impacts of these systems in other categories such as ozone depletion, acidification, eutrophication, smog formation, ecotoxicity, resource use, or human health impacts. 

\section{Results and discussion} \label{results}
 
This section compares and analyzes the environmental impact results for the autonomous, station-based, and dockless BSS. First, we analyze the three systems for the nominal scenario, defined by the values described in the life cycle inventory (Section \ref{inventory}). Then, we estimate the impacts of different hypotheses to account for the uncertainty in the assumptions made at this early technology state and identify the parameters that have the most impact on the environmental performance of autonomous BSS. Finally, we study the overall impact of the three types of BSS when accounting for the environmental impacts of the mobility modes substituted by the bike-sharing trips.

\subsection{Nominal scenario}
\begin{figure}[!htb]
    \centering
    \includegraphics[width=0.7\linewidth]{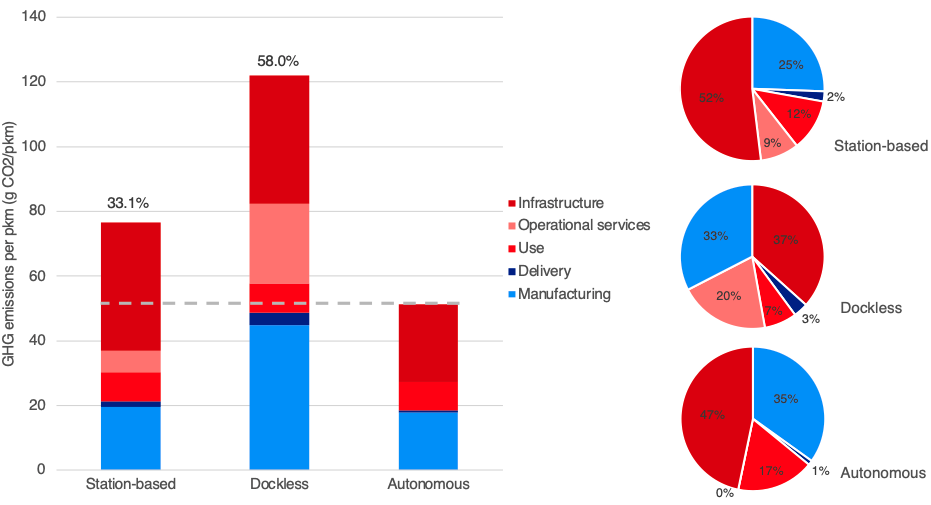}
    \caption{Nominal scenario GHG emissions for the different LCA components for the three systems considered: station-based, dockless, and autonomous BSSs. Left: Total GHG emissions in g $\mathrm{CO_2}$ equivalent per passenger kilometer traveled for each LCA component. Right: Share of emissions for each LCA component. The extended line represents the autonomous system, which is taken as a baseline. } 
    \label{fig:nominal}
\end{figure}

The results for the nominal scenario can be observed in Figure \ref{fig:nominal} and Table \ref{tb:nominal-results}. The results show that, with the assumptions considered in this study, an autonomous BSS would emit 51.23 g $\mathrm{CO_2}$/pkm; that is 33.1 \% less than the station-based system and 58.0 \% less than the dockless system which respectively emit 76.59 g $\mathrm{CO_2}$/pkm and 122.07 g $\mathrm{CO_2}$/pkm.

The difference is because, as the autonomous system is more efficient than the current BSS, it needs fewer bikes. Moreover, even if the need for infrastructure is higher on a per bicycle basis, double than in the station-based system in the nominal scenario (Section \ref{inventory}), it is lower overall. In addition, as bicycles can relocate by themselves, there is no need for rebalancing vans, which are one of the most significant contributors to the environmental impacts of the current BSS \cite{luo2019comparative}. Interestingly, even if the impact of manufacturing each autonomous bicycle is significantly higher than the station-based and dockless bicycles (826.9 kg $\mathrm{CO_2}$/bike compared to 188.0 kg$\mathrm{CO_2}$/bike ), due to the difference in fleet sizes, the autonomous system has the lowest total environmental cost of manufacturing the bicycles.

It is important to note that infrastructure is the main contributor to the environmental impact of the three systems, representing 52 \% for the station-based system, 37 \% for the dockless system, and 47 \% for the autonomous system. The environmental impacts of the shared e-bike systems in this study are higher than the one calculated by the ITF - 80 g $\mathrm{CO_2}$/pkm for dockless e-bikes - because they do not consider the impact of docking and charging stations within the system bounderies \cite{ITF}. This result highlights the relevance of considering charging infrastructure within the boundaries of the LCA analysis. 


Our results seem consistent with those reported by Luo et al. for station-based, and dockless BSS \cite{luo2019comparative}. Luo et al. estimate 65 g $\mathrm{CO_2}$/pkm for station-based systems and 116 g $\mathrm{CO_2}$/pkm for dockless systems considering conventional bikes. Since we model e-bikes, it was expected that our values would be higher but are still within a reasonable range being 15 \% higher for the station-based system and 5\% for the dockless system. However, a limitation that should be noted is that we are combining the vehicle usage data from \cite{sanchez2021simulation} which considers regular bicycles, but accounting for the environmental performance of e-bikes. Since e-bikes tend to have higher usage rates than regular bikes \cite{fyhri2015effects}, this study likely overestimates the negative impacts of the BSS.

\begin{table}[]
\centering
\scalebox{0.7}{
\begin{tabular}{llll} \hline
  & Station-based & Dockless & Autonomous \\ \hline
Lifetime [years] & 3 & 3 & 3 \\
Daily mileage [km] & 8.77 & 3.84 & 42.23 \\
Bicycle usage [trips/bike/day] & 2.5 & 1.1 & 8.8 \\
Electricity Use & USA & USA & USA \\
Weight [kg] & 30.8 & 30.8 & 40.4 \\
Battery capacity [kWh] & 0.48 & 0.48 & 0.49 \\
Additional components for autonomy [kg $\mathrm{CO_2}$] & - & - & 599.3 \\
Rebalancing [km/pkm] & 0.03 & 0.1 & - \\
Stations [station/pkm] & 4.38E-06 & - & 2.5E-06 \\ \hline
Vehicle and battery  manufacturing, assembly and disposal {[}g $\mathrm{CO_2}$/pkm{]} & 19.58 & 44.76 & 17.88 \\
Vehicle delivery at point of purchase {[}g $\mathrm{CO_2}$/pkm{]} & 1.73 & 3.96 & 0.52 \\
Vehicle use (including fuel production) {[}g $\mathrm{CO_2}$/pkm{]} & 8.88 & 8.88 & 8.88 \\
Operational services {[}g $\mathrm{CO_2}$/pkm{]} & 6.62 & 24.70 & - \\
Infrastructure {[}g $\mathrm{CO_2}$/pkm{]} & 39.77 & 39.77 & 23.94\\ \hline
\textbf{Total GHG emissions per pkm {[}g $\mathrm{CO_2}$/pkm{]}} & \textbf{76.59} & \textbf{122.07} & \textbf{51.23} \\ \hline
\end{tabular}}
    \caption{Summary of values considered for the nominal scenario analysis of the environmental impact of the three systems considered (station-based, dockless and autonomous BSS) and the corresponding results expressed in equivalent $\mathrm{CO_2}$ emissions per passenger kilometer.}
    \label{tb:nominal-results}
\end{table} 

\subsection{Sensitivity analyses}

Autonomous BSS being at such an early development stage, the characteristics and impacts of this technology have a high degree of uncertainty. However, this parameter variation and scenario analysis allow us to assess the impacts of variations in the modeling hypotheses, as well as to identify the parameters that have the most impact on the total environmental impact, consequently limiting the uncertainty of this study \cite{finnveden2009recent}.

\subsubsection{Vehicle lifetime}

\begin{figure}[!htb]
    \centering
    \includegraphics[width=1\linewidth]{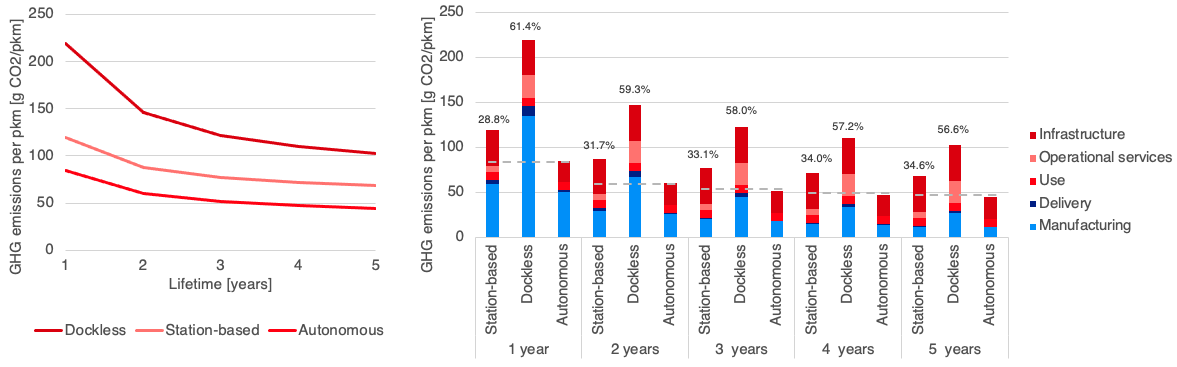}
    \caption{Life cycle GHG emissions of the three systems studied (station-based, dockless, and autonomous BSSs) expressed in equivalent $\mathrm{CO_2}$ emissions per passenger kilometer, as a function of variations in the vehicle lifetime assumption in the range of 1 to 5 years. For each scenario, the extended line represents the case of the autonomous system, which is taken as a baseline.} 
    \label{fig:lifetime}
\end{figure}

The impacts of varying the assumptions of vehicle lifetime from one to five years are shown in Figure \ref{fig:lifetime}. Manufacturing is the second-largest contributor to the environmental impacts of the three BSS (see Figure \ref{fig:nominal}); therefore, a variation in vehicle lifetime has a significant impact on environmental performance.  
If the lifetime is reduced from three to one year, the overall environmental impact increases by 55.67 \% for the station-based system, 79.81 \% for the dockless system, and 64.45 \% for the autonomous system. If the lifetime is extended from one to five years, instead, the overall impact decreases by 11.12 \% for the station-based system, 15.96 \% for the dockless system, and 13.14 \% for the autonomous system.

Results worsen significantly with shorter lifetimes, especially in the case of the dockless BSS due to the larger fleet sizes. In fact, with a lifetime of one year, the environmental impact of the dockless system would be even higher than for an ICE private car (see Figure \ref{fig:comparative}). While the autonomous BSS has the lowest emissions in all scenarios, the considerable variation in environmental impact indicates the importance of designing the vehicles to be durable and robust. 


\subsubsection{Vehicle utilization}

The results in the nominal scenario assume different vehicle utilization rates for the station-based, dockless, and autonomous BSS. These assumptions are based on previous studies that show that autonomous shared micro-mobility is expected to have levels of use well above the ones for current station-based and dockless systems \cite{sanchez2021simulation,kondor2021estimating}. However, since it is an important variable to study, we study what would happen if all the systems had the same vehicle utilization rates.

\begin{figure}[!htb]
    \centering
    \includegraphics[width=1\linewidth]{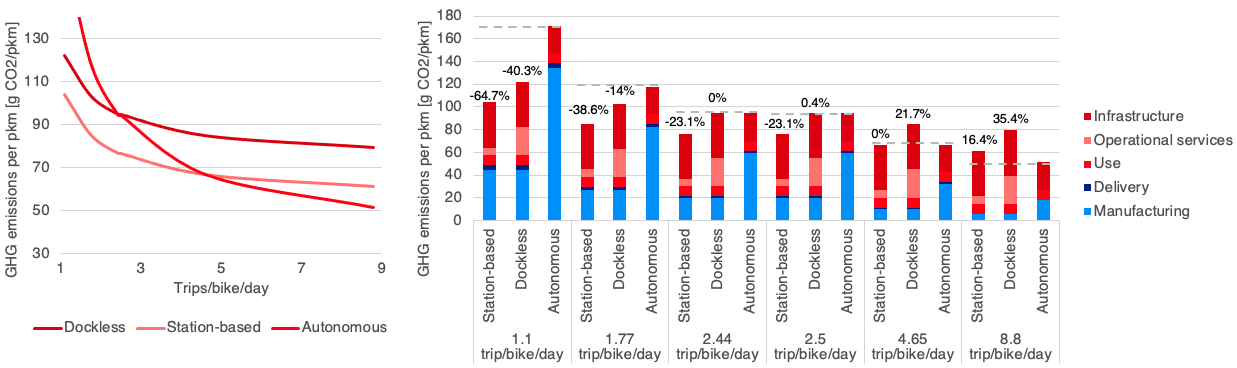}
    \caption{Life cycle GHG emissions of the three systems studied (station-based, dockless, and autonomous BSSs) expressed in equivalent $\mathrm{CO_2}$ emissions per passenger kilometer, as a function of the vehicle utilization rates in trips per bike and day. Values represented include 1.1 trip/bike/day (nominal scenario for the dockless system), 1.77 trip/bike/day (the value for the autonomous system equals the one for the dockless system), 2.5 trip/bike/day (nominal scenario for the station-based system), 4.65 trip/bike/day (the value for the autonomous system equals the one for the station-based system), and 8.8 trip/bike/day (nominal scenario for the autonomous system) \cite{sanchez2021simulation}. For each scenario, the extended line represents the case of the autonomous system, which is taken as a baseline.}
    \label{fig:tripsbikeday}
\end{figure}

As can be observed in Figure \ref{fig:tripsbikeday}, if autonomous bicycles were as inefficient in the number of trips per bike and day as current dockless systems, their environmental impact would be 64.7 \% and 40.3 \% higher than with current station-based and dockless systems BSS. As the emissions of manufacturing an autonomous bicycle are more than four times higher than for current shared e-bikes, it is not surprising the vehicle utilization needs to be higher to compensate the impact of each bicycle. 
However, with a vehicle utilization rate of 2.44 trips/bike/day, which is very close to the typical value for current station-based systems \cite{sanchez2020autonomous}, the autonomous system already outperforms the dockless system. For the autonomous system to be more sustainable than the station-based system, the bikes should make at least 4.65 trips/bike/day, and autonomous bicycles have been estimated to make 8.8 trips/bike/day \cite{sanchez2021simulation}. As the advantage further increases as bicycle utilization rates increase, with this vehicle use rate, the autonomous system would outperform the station-based system 16.4 \% and the dockless system by 35.4 \%, even if dockless systems and station-based systems had the same vehicle usage.

\subsubsection{Rebalancing}

\begin{figure}[!htb]
    \centering
    \includegraphics[width=0.85\linewidth]{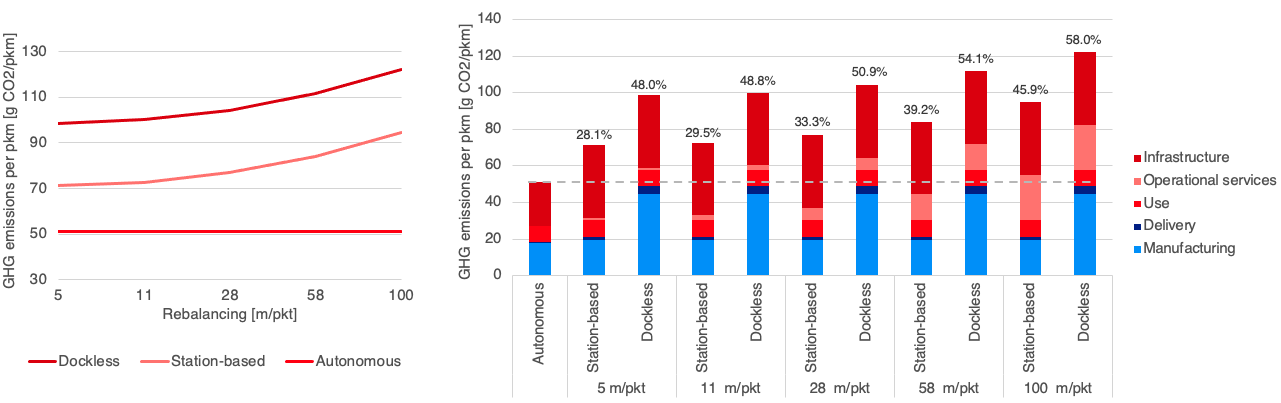}
    \caption{Life cycle GHG emissions of the three systems studied (station-based, dockless, and autonomous BSSs) expressed in equivalent $\mathrm{CO_2}$ emissions per passenger kilometer, as a function of the rebalancing need in meters of operational service needed per passenger kilometer traveled: 5 m/pkt, 11 m/pkt, 28 m/pkt, and 58 m/pkt (as in the optimistic, best-case, pessimistic and worst-case scenarios by Bortoli et al. \cite{de2021environmental}) and 100 m/pkt (as for shared e-bikes in \cite{ITF}). The extended line represents the nominal scenario for the autonomous system, which is taken as a baseline.} 
    \label{fig:rebalancing}
\end{figure}

The impact of different assumptions in the rebalancing need of the station-based and dockless systems can be found in Figure \ref{fig:rebalancing}. The nominal scenario considered different rebalancing needs for dockless and station-based systems assuming that dockless e-bikes need to be transported for both rebalancing and charging. The results show that assuming the same rebalancing need for station-based and dockless systems reduces the difference in their environmental impacts. However, the station-based system remains a more efficient option. 
Most importantly, this analysis shows that even in the most optimistic rebalancing scenario for station-based and dockless systems, their environmental impact remains 28.1 \% and 48.0 \% higher than the autonomous systems.

\subsubsection{Power mix and electrification}

\begin{figure}[!htb]
    \centering
    \includegraphics[width=0.85\linewidth]{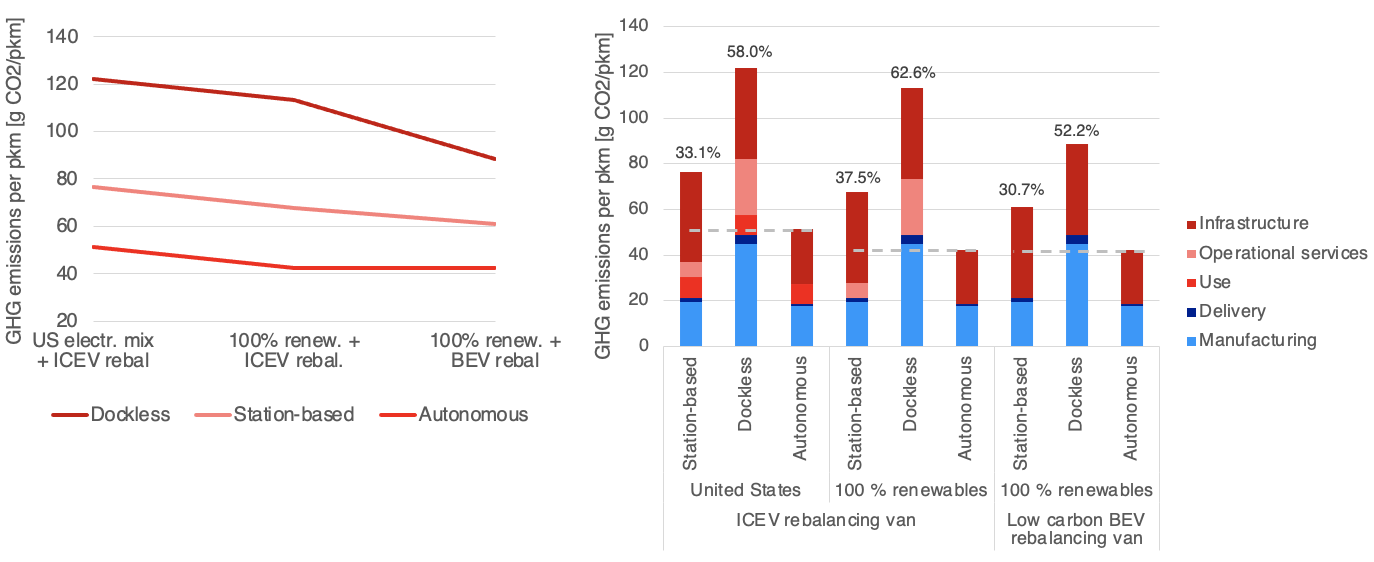}
    \caption{Life cycle GHG emissions of the three systems studied (station-based, dockless, and autonomous BSSs) expressed in equivalent $\mathrm{CO_2}$ emissions per passenger kilometer, considering two scenarios: (1) ICEV rebalancing vans combined with the electricity mix from the United States and a zero-carbon grid and (2) low-carbon BEV rebalancing vans combined with a zero carbon grid. For each scenario, the extended line represents the case of the autonomous system, which is taken as a baseline.} 
    \label{fig:electricity}
\end{figure}

As stated in Section \ref{inventory}, the electricity mix considered in this article is the one of the United States in 2020 \cite{IEA}. However, the electricity grid will likely be largely or fully decarbonized for when autonomous bicycles are implemented. Similarly, we could expect rebalancing vans to be battery electric vehicles (BEVs). We have analyzed the impacts of these assumptions in Figure \ref{fig:electricity}, where it can be observed that decarbonizing the grid impacts the use phase of the three systems. The overall lifecycle emissions are reduced by 11.6 \% for the station-based system, 7.3 \% for the dockless system, and 17.34 \% for the autonomous system. 
If the decarbonization of the grid is combined with switching ICEV rebalancing vans to low-carbon BEVs, the impact of the station-based and dockless systems can be further reduced by 9.8 \% and 21.82 \%. Overall, while these improvements reduce the impact of rebalancing, the autonomous system still performs significantly better than the current BSS in all the scenarios considered.

\subsubsection{Autonomous shared bikes}

\begin{figure}[!htb]
    \centering
    \includegraphics[width=0.7\linewidth]{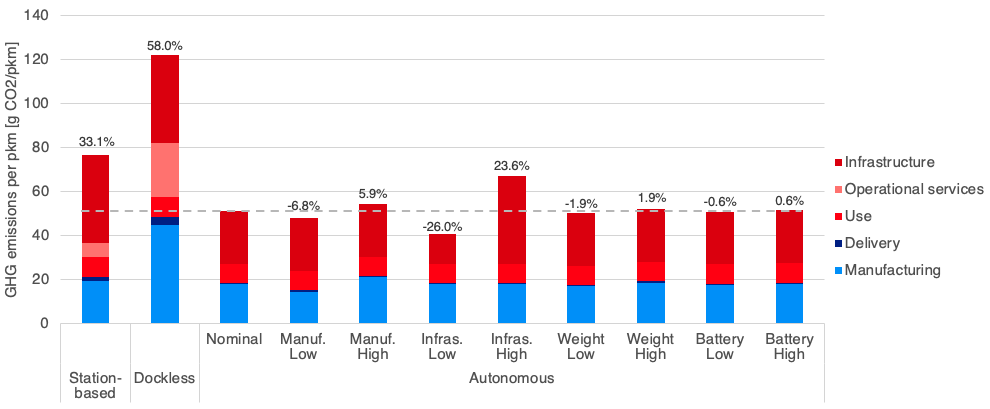}
    \caption{Life cycle GHG emissions of station-based, dockless, and autonomous BSSs considering different scenarios for the autonomous system: Impact of components for autonomous driving 25 \% less (Manuf. Low) and 25 \% more (Manuf. High) than in the nominal scenario. Same number of stations as in the station-based in a per bicycle basis (Infras.Low ) and same total number of stations than in the station-based system (Infras. High). Same weight as e-bike weight reported by \cite{ITF} (Weight Low), same as the PEV \cite{PEV} (Weight High). Battery capacity 25 \% less (Battery Low) and 25 \% more (Battery High) than in the nominal scenario. The extended line represents the nominal scenario for the autonomous system, which is taken as a baseline.}
    \label{fig:autonomous}
\end{figure}

This last section of the sensitivity analysis aims to account for the uncertainty in the assumptions related to autonomous bicycles. As can be observed in Figure \ref{fig:autonomous}, the assumptions in weight and battery capacity have a relatively small impact: $\pm 1.9 \%$ for the variations in weight and $\pm 0.6 \%$  for the variations in battery capacity. The technology that would support autonomous driving, instead, has a higher impact on the life-cycle emissions: a 25 \% decrease from the nominal scenario value of 599.3  kg $\mathrm{CO_2}$ would decrease the overall emissions by 6.8 \%, and an increase of 25 \% would drive an increase of 5.9 \%.

Once again, the infrastructure stands out as a very relevant factor. If the total need for infrastructure was the same as for station-based on a per bicycle basis - instead of being twice as high as the hypothesis for the nominal scenario - the impact of an autonomous BSS would be 26 \% lower than for the nominal scenario. In that case, the impact of the autonomous system would be 46.9 \% below the impact of a station-based system and 66.7 \% below the dockless system. Conversely, if the total need for infrastructure would be the same as for station-based bicycles, being approximately 3.5 times higher on a per bicycle basis, the autonomous BSS could have 23.6 \% more environmental impact than in the nominal scenario. In this case, the autonomous system would still have the best performance but by a smaller percentage, with 12.4 \% lower emissions than the station-based system and 45.1 \% lower emissions than the dockless system.

We would argue that two metrics that have the most significant influence on environmental performance -i.e., the need for infrastructure and the technology needed for autonomous driving - will present essential trade-offs in the future. Overall, there will be a trade-off between increasing battery capacity or increasing the infrastructure, and the same will happen for in-vehicle technology and embedding the intelligence in the city. This study considers that the infrastructure would be exclusive for BSS but, in a future in which autonomous and electric vehicles would be mainstream, the infrastructure could be shared to minimize costs and environmental impact. 

\subsection{Transportation mode substitution}\label{mode-substitution}

\begin{figure}[!htb]
    \centering
    \includegraphics[width=0.7\linewidth]{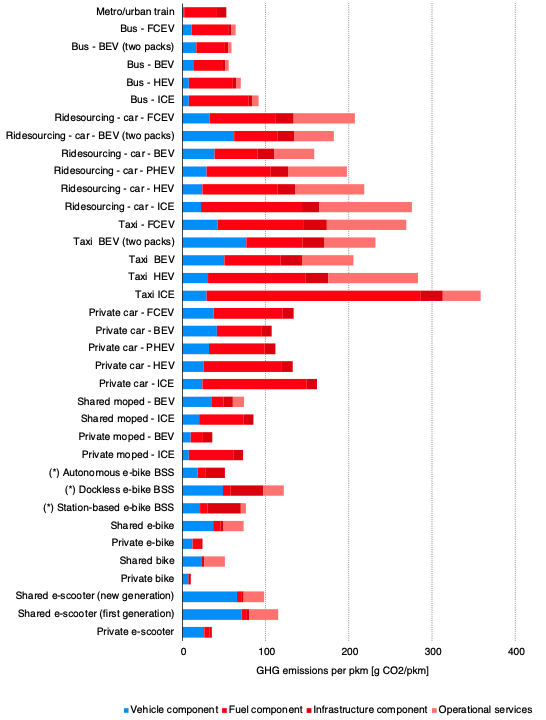}
    \caption{
     Life cycle GHG emissions of the station-based, dockless and autonomous BSS from this study (*) expressed in equivalent $\mathrm{CO_2}$ emissions per passenger kilometer for each LCA component, compared to other new and traditional mobility modes, adapted from \cite{ITF}.} 
    \label{fig:comparative}
\end{figure}

The results considered until this section reflect the environmental impacts of station-based, dockless and autonomous BSS. However, these systems' overall environmental impact will depend on the mobility modes displaced by the bike-sharing trips. In order to conduct this analysis, it is critical to have a consistent set of modeling assumptions for the new mobility mode -the BSS in this case- and the displaced traditional modes. We achieved this by also using the same assessment tool to calculate the environmental impacts of the displaced mobility modes, with the same energy mix assumptions as in Section \ref{inventory} \cite{ITF}. The environmental impacts of all the mobility modes, including the station-based, dockless and autonomous BSS calculated from this study, can be observed in Figure \ref{fig:comparative}. These results show that the environmental impact of an autonomous BSS per passenger kilometer traveled is above private micro-mobility modes but below any other type of shared micro-mobility modes. In addition, the impact of autonomous shared bikes is also below private ICE mopeds, public transit, and all private car, taxi, or ride-sourcing services.

In order to calculate the overall impact, these results have to be combined with mode shift data. Mode shift is very dependent on location, time, and travel context, being very much trip-specific \cite{li2021high}. To account for this variability, we have studied a range of reported mode shifts from different countries gathered by Teixeira et al. \cite{teixeira2021empirical}. Moreover, we have considered two vehicle electrification scenarios, with values are based on Figure \ref{fig:comparative}. Scenario 1 is the most favorable scenario, which considers private cars and taxis to be ICEV (162 g $\mathrm{CO_2}$/pkm and 358 g $\mathrm{CO_2}$/pkm, respectively) and ICEV buses as public transit (91 g $\mathrm{CO_2}$/pkm). Scenario 2 reflects the least favourable scenario by considering lower-emitting versions; it considers BEV cars (108 g $\mathrm{CO_2}$/pkm), BEV taxis (206 g $\mathrm{CO_2}$/pkm) and metro/urban train as public transit option (52 g $\mathrm{CO_2}$/pkm).

As can be observed in Table \ref{mode-substitution}, according to these hypotheses, the station-based BSS only has a positive environmental impact with the mode shares of Brisbane, Hangzhou, and Paris in Scenario 1. It has an overall negative environmental impact in the rest of the cities for Scenario 1 and all cities for Scenario 2. The dockless BSS has an overall negative environmental impact in most scenarios. Lastly, the autonomous system provides an improvement in all cases except Dublin in Scenario 1, while in Scenario 2 it only improves the environmental performance in the cases of Brisbane, Hangzhou, and Paris. These results are mainly derived from combination of low substitution rates from cars (3-23\%), a high substitution rate from walking (19-46 \%) and private bicycles (7-28 \%), and the generation of new trips (1-10 \%). With these mode shifts, the environmental impact of BSS should be below 63 g $\mathrm{CO_2}$/pkm to have a positive impact in more than half of the cities in Scenario 1 and below 40 g $\mathrm{CO_2}$/pkm for Scenario 2. In fact, in order to have a positive impact in the most restraining case, which is Dublin in Scenario 2, the impact should be below 21 g $\mathrm{CO_2}$/pkm, which is below the impact of private e-bikes. 

On one hand, in the worst-case mode shift that is Dublin, only 3 \% of the trips would have been otherwise made by car, 31 \% in public transit, 54 \% walking, and 12 \% with their own bike. For that mode shift, the three types of BSS would have a negative environmental impact with increases of 117.4 \% for the station-based system, 246.5 \% for the dockless system, and 45.38 \% for the autonomous system in Scenario 1, and 255.7 \%, 467.0 \%, and 137.92 \%, respectively, for Scenario 2.
Paris, on the other hand, has the best-case mode shift: 8 \% of the trips would have otherwise been done by car, 5 \% by taxi, 65 \% in public transit, and 20 \% walking. For this mode shift, the station-based system would improve the environmental performance by 12.19 \% in Scenario 2 and worsen it by 44.5 \% in Scenario 2. The dockless system would have a negative environmental impact even for this best-case scenario, increasing the environmental impacts 35.2 \% in Scenario 1 and 130.3 \% in Scenario 2. Finally, the autonomous system would positively impact both scenarios with a 43.28 \% decrease in Scenario 1 and 3.36 \% in Scenario 2.

It is important to note that in this study, we consider the mode shift of regular bikes but account for the environmental performance of e-bikes. Since the mode-shift created by e-bikes could be different by, for instance, substituting longer trips that are more frequently done by car or public transit, this study may overestimate the negative impacts of the BSS. 

All in all,  while the autonomous system provides improved performance over current BSS systems, its benefits will ultimately be subject to the mod-shift that it generates. It is a common challenge to station-based and dockless BSS to drive a shift from private cars and some studies point towards the need to combine bike-sharing with public transport as a solution to replace more car trips \cite{teixeira2021empirical,shaheen2013public}. This study shows that as private vehicles, taxis, and public transit electrify and become less pollutant, improving the overall environmental performance will be more challenging, and autonomous BSS will also need to become more efficient and focus on replacing almost exclusively more polluting mobility modes.

\begin{table}[]
\centering
\scalebox{0.65}{
\begin{tabular}{lllllllllllll}
\hline
 &\multicolumn{6}{c}{Substituted   mode share}  & \multicolumn{6}{c}{Environmental impact   of the BSS} \\ 
 & \multicolumn{6}{c}{}& \multicolumn{2}{c}{Station-based} & \multicolumn{2}{c}{Dockless} & \multicolumn{2}{c}{Autonomous} \\ \hline
 & Car/motorc. & Taxi & PT & Walking & Own bike & New Trip & Scen.1 & Scen.2 & Scen.1 & Scen.2 & Scen.1 & Scen.2 \\\hline
Birsbane & 21\% & 3\% & 43\% & 23\% & 8\% & 1\% & -9.46\% & 46.85\% & 44.31\% & 134.05\% & -39.45\% & -1.79\% \\
Melbourne & 19\% & 2\% & 41\% & 27\% & 9\% & 1\% & 0.52\% & 62.55\% & 60.21\% & 159.07\% & -32.77\% & 8.71\% \\
Hangzhou & 8\% & 5\% & 59\% & 19\% & 10\% &  & -11.47\% & 48.57\% & 41.09\% & 136.79\% & -40.79\% & -0.64\% \\
Nanjing & 15\% & 4\% & 19\% & 47\% & 15\% & & 30.82\% & 107.72\% & 108.50\% & 231.06\% & -12.51\% & 38.92\% \\
Barcelona & 10\% &  & 51\% & 26\% & 6\% & & 19.97\% & 98.74\% & 91.21\% & 216.75\% & -19.77\% & 32.92\% \\
Dublin & 3\% & & 31\% & 54\% & 12\% & & 117.37\% & 255.74\% & 246.45\% & 466.98\% & 45.38\% & 137.92\% \\
Dublin & 20\% &  & 35\% & 46\% & &  & 18.94\% & 92.07\% & 89.57\% & 206.12\% & -20.45\% & 28.45\% \\
Lyon & 7\% & & 50\% & 37\% & 4\% & 2\% & 35.32\% & 126.78\% & 115.68\% & 261.44\% & -9.50\% & 51.67\% \\
London & 1\% & 4\% & 61\% & 29\% & 5\% & 1\% & 6.58\% & 83.44\% & 69.88\% & 192.37\% & -28.72\% & 22.68\% \\
London & 2\% & 4\% & 58\% & 26\% & 8\% & 4\% & 10.69\% & 88.85\% & 76.42\% & 200.99\% & -25.97\% & 26.30\% \\
UK & 23\% & 3\% & 22\% & 31\% & 9\% & 10\% & 20.97\% & 91.58\% & 92.80\% & 205.34\% & -19.10\% & 28.13\% \\
UK & 13\% & 3\% & 28\% & 33\% & 12\% & 10\% & 41.73\% & 128.44\% & 125.89\% & 264.09\% & -5.21\% & 52.78\% \\
UK & 10\% & 4\% & 27\% & 42\% & 7\% & 7\% & 45.29\% & 138.65\% & 131.56\% & 280.36\% & -2.83\% & 59.61\% \\
Paris & 8\% & 5\% & 65\% & 20\% & &  & -15.19\% & 44.49\% & 35.17\% & 130.29\% & -43.28\% & -3.36\% \\
Minnesota & 19\% & 3\% & 20\% & 38\% & 8\% & 9\% & 36.48\% & 117.09\% & 117.52\% & 246.00\% & -8.72\% & 45.19\% \\
Montreal & 2\% & 8\% & 33\% & 25\% & 28\% & 4\% & 19.19\% & 95.84\% & 89.96\% & 212.12\% & -20.29\% & 30.97\% \\
Montreal & 8\% & & 50\% & 18\% & 24\% &  & 22.08\% & 96.83\% & 94.57\% & 213.71\% & -18.35\% & 31.64\% \\
Montreal & 10\% & 6\% & 41\% & 21\% & 22\% & & -2.95\% & 58.43\% & 54.68\% & 152.50\% & -35.09\% & 5.96\% \\
Washington & 7\% & 6\% & 45\% & 31\% & 6\% & 4\% & 6.20\% & 78.97\% & 69.26\% & 185.24\% & -28.98\% & 19.69\%\\ \hline
\end{tabular}}
    \caption{Share of mobility modes displaced by BSS in different countries \cite{teixeira2021empirical} and the corresponding impact of the introduction of station-based, dockless and autonomous BSS in as the relative percentage of increase or decrease in GHG emissions. Scenario 1 considers cars and taxis to be ICEV and public transit (PT) to be ICEV buses. Scenario 2 considers cars and taxis to be BEV and PT to be a metro/urban train (values from Fig.\ref{fig:comparative}). }
    \label{tb:mode-substituton}
\end{table}

\section{Conclusions}

Over the last fifteen years, cities worldwide have implemented BSS, intending to reduce the environmental emissions from transportation. However, it was not until 2019 that their environmental impacts started to be assessed. Recently, autonomous bicycles have been proposed as a more convenient and efficient alternative to current station-based and dockless systems, but the corresponding environmental implications remained unresearched. With the climate change issues being more pressing than ever before, it is critical to understand the environmental impacts of new mobility modes before they are deployed. 

An early technology stage is when there is a greater opportunity to influence their design and, ultimately, their environmental impacts once they are deployed. Consequently, this study compares the environmental impacts of autonomous bicycles compared with current station-based and dockless systems under different sets of modeling hypotheses and mode-shift scenarios. The results indicate that introducing autonomous driving technology into BSS can improve their environmental performance, reducing the environmental impact per passenger kilometer traveled by 33.1 \% compared to station-based systems and 58.0 \% compared to dockless systems for the nominal scenario. However, the sensitivity analysis shows that the environmental impact of autonomous shared bicycles will mainly depend on the levels of use and the need for infrastructure. In addition, other design variables such as increasing vehicle lifetime and reducing the components needed for autonomous driving will also significantly reduce the environmental impacts of autonomous BSS.
The improvement in the environmental performance of autonomous shared bikes over current systems becomes especially significant when accounting for the displaced mobility modes. Currently, most bike-share trips currently substitute walking, private bikes, or new trips induced by the BSS. Therefore, it will be critical to consider that, as other mobility modes decarbonize and become more efficient, autonomous BSS should focus on targeting mode replacement from more polluting modes.

As a final note, we would like to remind that, while LCA is a valuable tool for understanding the environmental impacts of products and services, methodological choices heavily influence the results. Conducing an environmental impact analysis of an emerging technology poses an even more significant challenge due to the uncertainties derived from the lack of real-world data and, while we have tried to limit the uncertainties by focusing the efforts on broader scenario analysis, we advise readers to keep in mind that environmental impacts are very case-specific. 

 \bibliographystyle{elsarticle-num} 
 \bibliography{cas-refs}





\end{document}